\documentclass[runningheads]{llncs}
\usepackage[T1]{fontenc}
\usepackage{mathtools}
\usepackage{graphicx}

\usepackage{amsmath}
\usepackage{bm}

\usepackage{framed,multirow}

\usepackage{wrapfig} % Required for text to wrap around figures and tables
\usepackage[labelfont=bf]{caption} % Make figure numbering in captions bold

\usepackage[T1]{fontenc}
\usepackage{multirow}
\usepackage{graphicx}
\usepackage{amssymb}
\usepackage{amsfonts}
\usepackage{booktabs}
\usepackage{array}

\usepackage{xcolor,soul}
\sethlcolor{lightgray}

\newcolumntype{L}[1]{>{\raggedright\let\newline\\\arraybackslash\hspace{0pt}}m{#1}}
\newcolumntype{C}[1]{>{\centering\let\newline\\\arraybackslash\hspace{0pt}}m{#1}}
\newcolumntype{R}[1]{>{\raggedleft\let\newline\\\arraybackslash\hspace{0pt}}m{#1}}

\makeatletter
\newcommand{\thickhline}{%
    \noalign {\ifnum 0=`}\fi \hrule height 1pt
    \futurelet \reserved@a \@xhline
}
\newcolumntype{"}{@{\hskip\tabcolsep\vrule width 1pt\hskip\tabcolsep}}
\makeatother

\begin{document}

\title{Direct segmentation of brain white matter tracts in diffusion MRI}
\titlerunning{White matter tract segmentation}

\author{Hamza Kebiri\inst{1,2}, and Ali Gholipour\inst{1}, Meritxell Bach Cuadra\inst{1,2}, \\ Davood Karimi\inst{1}}
% index{Karimi, Davood}

\institute{Computational Radiology Laboratory (CRL), Department of Radiology, Boston Children's Hospital, and Harvard Medical School, USA \and Department of Radiology, Lausanne University Hospital (CHUV) and University of Lausanne (UNIL), Lausanne, Switzerland}%
%

% \author{Anonymous}
% % \author{First Author\inst{1}\orcidID{0000-1111-2222-3333} \and
% % Second Author\inst{2,3}\orcidID{1111-2222-3333-4444} \and
% % Third Author\inst{3}\orcidID{2222--3333-4444-5555}}
% %
% % \authorrunning{F. Author et al.}
% \institute{Anonymous Organization \\
% \email{***@****.***}}
%
\maketitle              % typeset the header of the contribution

\begin{abstract}

The brain white matter consists of a set of tracts that connect distinct regions of the brain. Segmentation of these tracts is often needed for clinical and research studies. Diffusion-weighted MRI offers unique contrast to delineate these tracts. However, existing segmentation methods rely on intermediate computations such as tractography or estimation of fiber orientation density. These intermediate computations, in turn, entail complex computations that can result in unnecessary errors. Moreover, these intermediate computations often require dense multi-shell measurements that are unavailable in many clinical and research applications. As a result, current methods suffer from low accuracy and poor generalizability. Here, we propose a new deep learning method that segments these tracts directly from the diffusion MRI data, thereby sidestepping the intermediate computation errors. Our experiments show that this method can achieve segmentation accuracy that is on par with the state of the art methods (mean Dice Similarity Coefficient of 0.826). Compared with the state of the art, our method offers far superior generalizability to undersampled data that are typical of clinical studies and to data obtained with different acquisition protocols. Moreover, we propose a new method for detecting inaccurate segmentations and show that it is more accurate than standard methods that are based on estimation uncertainty quantification. The new methods can serve many critically important clinical and scientific applications that require accurate and reliable non-invasive segmentation of white matter tracts.


\end{abstract}

% Novelty Statement: Non-invasive segmentation of brain white matter tracts is enabled with Diffusion MRI. However, existing methods have low accuracy and poor generalizability because they depend on complex intermediate computations. We propose a new method that achieves superior accuracy by segmenting the tracts directly from the data. We also propose a novel method for detecting inaccurate segmentations.

\section{Introduction}

The brain white matter is organized into a set of distinct tracts. These tracts are bundles of myelinated axons that connect different brain regions such as the cerebral cortex and the deep gray matter. Although they are tightly packed and often cross one another, each tract has an entirely different function and connects different regions of the brain \cite{wakana2004fiber,wycoco2013white}. Accurate segmentation of these tracts is needed in clinical studies and medical research. For example, in surgical planning one needs to know the precise extent of the individual tracts in order to assess the risk of damage to specific neurocognitive functions that may result from surgical removal of brain tissue. As another prominent example, changes in the micro-structural properties of different tracts is commonly used in studying brain development and disorders.

Magnetic resonance imaging (MRI) is the modality of choice for non-invasive assessment of white matter tracts in vivo. Although some of the tracts may be identifiable on T1, T2, or FLAIR images \cite{wycoco2013white}, accurate segmentation of most tracts is only possible with diffusion MRI. Individual tracts may be extracted from whole-brain tractograms by specifying inclusion and exclusion regions of interest (ROIs). This process, which is usually referred to as ``virtual dissection'', is time-consuming, subjective, and it has low reproducibility \cite{schilling2021tractography}. Some prior works have aimed at automating the virtual dissection process by learning to compute the inclusion/exclusion ROIs \cite{suarez2012automated,yendiki2011automated}. It is also possible to extract the tracts from a whole-brain tractogram by grouping similar streamlines using a clustering approach. This can be done by comparing individual streamlines with a predefined set of tracts in an atlas \cite{garyfallidis2018recognition,labra2017fast}. Some techniques additionally take into account the location of the streamlines relative to anatomical landmarks in the brain \cite{siless2018anatomicuts,siless2020registration}. Tractography-based methods are inherently limited by the errors in streamline tractography \cite{maier2017challenge}. To avoid these errors, some methods segment the tracts on diffusion tensor or fiber orientation images, thereby avoiding the tractography. Some of the segmentation techniques that have been explored in the past include Markov Random Fields \cite{bazin2011direct}, k-nearest neighbors technique \cite{ratnarajah2014multi}, template matching \cite{eckstein2009active}, and more recently deep learning \cite{dong2019multimodality,wasserthal2018tractseg}. However, none of these intermediate parameters (e.g., the diffusion tensor) have an unambiguous biophysical meaning and their computation entails unavoidable estimation errors. Moreover, the intermediate computations for most existing methods assume availability of dense multi-shell diffusion MRI measurements, which are not acquired in many clinical and research applications. As a result, existing methods have low accuracy and limited generalizability when applied to typical clinical scans.

In this work, we develop and validate a new method that segments white matter tracts directly from the diffusion MRI data. The new method does not require tractography or computation of other intermediate parameters. Moreover, we present a simple but effective technique for detecting less accurate segmentations. We show that the new methods achieve superior accuracy and generalizability compared with the existing methods.

\section{Materials and methods}

\subsection{Segmentation model}

\begin{figure}[!htb]
\centering
\includegraphics[width=1.0\textwidth]{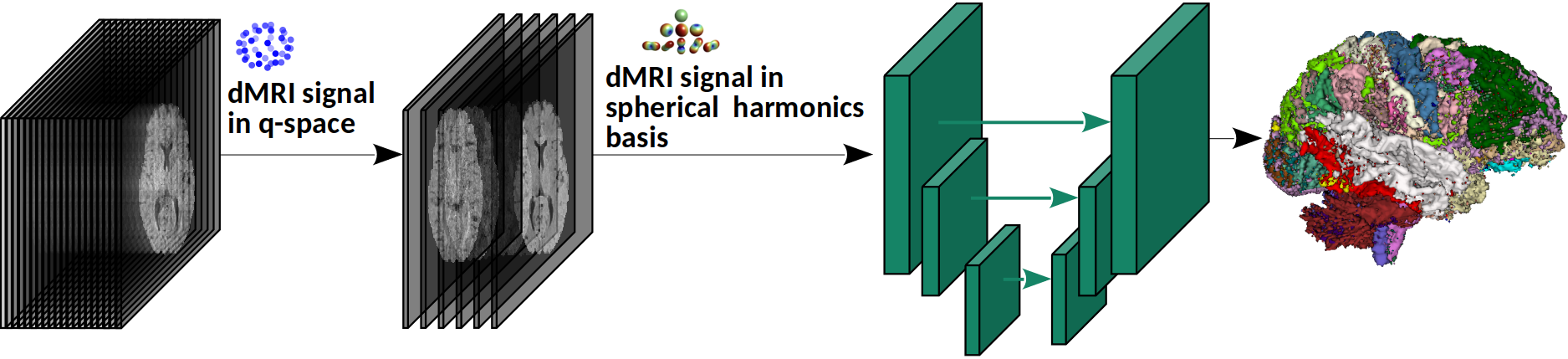}
\caption{\small{Overview of the proposed tract segmentation method.}}
\label{fig:segmentation_method}
\end{figure}

Our method, shown schematically in Figure \ref{fig:segmentation_method}, is based on a fully convolutional network (FCN). The network architecture is similar to nnU-Net (we refer to \cite{isensee2021nnu} for the details of the architecture). Our method predicts tract segmentations directly from the diffusion MRI data. To enhance the generalizability of the method and to enable it to work with scans acquired using different gradient tables (i.e., different gradient strengths and/or different gradient directions): (i) We train the model with measurements that are typically acquired for diffusion tensor imaging (DTI). DTI-style scans include single-shell measurements at a b-value of around 750-1200 s/mm$^{2}$ \cite{jones1999optimal}. They are the most common acquisition in clinical and research applications. We normalize these measurements by a non-weighted (b=0) measurement. (ii) We project the normalized data onto a fixed spherical harmonics (SH) basis. We use SH bases formulations of \cite{tournier2007robust} with an order 2, which results in 6 SH coefficients regardless of the number of measurements. We use these 6 coefficient maps as input to the FCN.

Our approach of using the data as the model input has three advantages:

(1) It eliminates the need to compute intermediate parameters (e.g., fiber orientation distribution or tractogram), thereby avoiding the associated computational errors \cite{schilling2019limits,schilling2019challenges}. If the goal is tract segmentation, there is no need to incur those errors by going through intermediate computations.

(2) It improves the generalizability of the method with respect to different acquisition schemes. If, for example, the input is the tractogram, the tract segmentation results can be significantly influenced by the tractography method that is used to compute the tractogram. Moreover, computation of intermediate parameters may demand especial measurement schemes that may be unavailable at test time. For example, methods that are based on fiber orientation distribution typically require high angular resolution measurements, which can result in a loss of accuracy if such measurements are not available \cite{descoteaux2009high,wasserthal2018tractseg}.

(3) It offers a highly effective data augmentation method during both training and test/inference. Data augmentation during training improves the training of large deep learning models with limited data. It is especially common in applications such as medical imaging where labeled data are costly to obtain. Test-time data augmentation, on the other hand, can be used to improve prediction accuracy and also to estimate prediction uncertainty \cite{abdar2021review,moshkov2020test,nalepa2019training}. Our train- and test-time data augmentation strategies are explained below.

Let us denote the set of b0-normalized measurements in a scan with $\{ x(q_i) \}_{i=1}^m$, where $q_i$ is the unit vector indicating the gradient direction for the $i^{\text{th}}$ measurement. During training, in each iteration we select a subset of size 6-12 from the $m$ measurements $\{ x(q_j) \}_{j \in S \subseteq \{1, \dots , m \} }$, chosen uniformly at random without replacement. We select these measurements such that the gradient directions for each pair of measurements are far apart in the q space, using an approach similar to \cite{jones1999optimal,skare2000condition}. We use the selected measurement subset (after projecting onto the SH basis) as input to the model. This can act as a highly effective and computationally-efficient data augmentation strategy as it presents a different view of the input to the model in each training iteration.

\begin{figure}[!htb]
\centering
\includegraphics[width=1.0\textwidth]{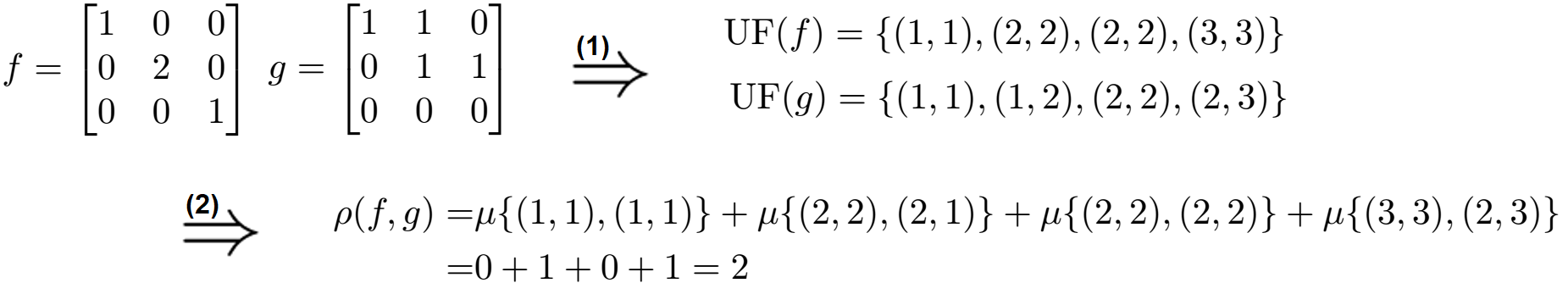}
\caption{\small{An illustration of the extension of the Wasserstein Distance to multi-dimensional signals, proposed in \cite{werman1985distance}.} }
\label{fig:EMD_computation}
\end{figure}

During inference, we use $n$ different measurement subsets, selected similarly as in training described above, to predict $n$ different segmentations. Let us denote the segmentation probability map for a specific tract with each of these measurement subsets as $\{ y_k \}_{k=1}^n$. We compute the voxel-wise average of these predictions to obtain a final segmentation prediction, which we denote with $\bar{y}$. Furthermore, we can compute a measure of disagreement between these $n$ predictions to estimate segmentation uncertainty. Disagreement between segmentation predictions is usually quantified using metrics of volume overlap or surface distance \cite{taha2015metrics}. Each of these metrics quantifies the segmentation error from a narrow perspective. Furthermore, these metrics discard the probability information by binarizing the segmentations. Recent segmentation uncertainty quantification methods have also followed a purely voxel-wise approach \cite{mehrtash2020confidence,wang2019aleatoric}, which ignores the spatial distribution of the segmentation probabilities. To characterize the disagreement in a way that accounts for the complete probability distribution of the predicted segmentations, we use a method based on the Wasserstein Distance, also known as earth mover's distance (EMD) \cite{rubner2000earth}. Given two probability distributions $p$ and $q$ defined on the same metric space, this distance is defined as $\mathtt{EMD}(p,q)= \inf_{\gamma \in \Gamma(p,q)} \mathbf{E}_{(x,y) \sim \gamma} d(x,y)$, where $d$ is a distance measure and $\Gamma(p,q)$ is the set of joint probability distributions whose marginals are equal to $p$ and $q$. Intuitively, if $p$ and $q$ are considered as two piles of earth, EMD is the cost of turning one into the other. Although EMD can be easily quantified for scalar variables, to the best of our knowledge there are no methods for computing EMD for probability distributions in ${\rm I\!R}^2$ or ${\rm I\!R}^3$. Here, we adopt an approximation that was originally proposed in \cite{werman1985distance} for comparing multi-dimensional histograms. We demonstrate this computation for a simple $3 \times 3$ histogram in Figure \ref{fig:EMD_computation}. Given a pair of multi-dimensional histograms (or probability distributions), the method first unfolds the histograms as shown in the example in Figure \ref{fig:EMD_computation} and finds a minimum distance pairing between the two. The distance between the two histograms is defined as the sum of the pair-wise distances in the pairing.

Based on this approximation, we compute the EMD between two segmentation probability maps in $\mathbb{R}^3$ as $\mathtt{EMD}(p,q)= \sum_{t=0}^1 d \big( P(t), Q(t) \big)$, where $P$ is the cumulative sum of unfolded $p$ as shown in Figure \ref{fig:EMD_computation} and the same for $Q$, and $d$ computes the $\ell_2$ distance between the paired $P$ and $Q$. This computation requires that the two inputs have the same mass, which we satisfy by normalizing the segmentations to have a unit sum. Furthermore, to reduce the computation time, we reduce the size of the segmentation volumes by a factor of 4 in each dimension via cubic interpolation. Given the set of $n$ segmentation predictions computed as explained above, we estimate the segmentation uncertainty as $u= \frac{1}{n} \sum_k \mathtt{EMD}(y_k, \bar{y})$.

\subsection{Implementation details}

The segmentation network was implemented in TensorFlow 1.6 and run on an NVIDIA GeForce GTX 1080 GPU on a Linux machine with 64 GB of memory and 20 CPU cores. The network takes 3D patches of size $96^3$ voxels as input and estimate the tract segmentation map for that patch. The network input has 6 channels as described above. The network output has 41 channels for the 41 tracts considered in this work. A complete description of these tracts can be found in \cite{wasserthal2018tractseg}. We merged the left and right sections of bilateral tracts, such as arcuate fasciculus, into one label. We trained the network to maximize the Dice similarity coefficient (DSC) between the predicted and ground-truth segmentation of the tracts using Adam \cite{kingma2014} with a batch size of 1 and a learning rate of $10^{-4}$, which was reduced by half if after a training epoch the validation loss did not decrease. We compare our method with TractSeg \cite{wasserthal2018tractseg}. TractSeg was shown to be vastly superior to many tractography dissection methods \cite{wasserthal2018tractseg}. Therefore, we do not compare with those methods.

\section{Experimental Results and Discussion}

We applied the method on 105 subjects in the Human Connectome Project study \cite{glasser2013minimal,van2013wu}. Manual segmentations of 72 tracts for these subjects are publicly available \cite{wasserthal2018tractseg}. We followed a five-fold cross-validation approach, each time leaving 21 subjects for test and training on the remaining 84 subjects. Table \ref{table:segmentation_accuracy_table} summarizes the performance of the proposed method and TractSeg. We report DSC, 95 percentile of the Hausdorff Distance (HD95), and average symmetric surface distance (ASSD). In addition to TractSeg, we compare our method with atlas-based segmentation (MAS), whereby 20 training images are registered to the test subject and the registration transforms are used to warp the segmentation labels from the training images to the test image. Voxel-wise averaging is then used to estimate the segmentations for the test image. We implemented this in two ways: MAS-FA, where we computed the registrations based on fractional anisotropy (FA) images using ANTS \cite{avants2009advanced}, and MAS-FOD, where we computed the registrations based on fiber orientation density images using mrregister \cite{raffelt2011symmetric}.

\begin{table*}[!htb]
\centering
\small
 \caption{\small{Segmentation performance of different methods. Asterisks denote significantly better results at $p=0.01$.}}
 \label{table:segmentation_accuracy_table}
\begin{tabular}{L{3.0cm}  L{1.8cm} | C{2.20cm} C{2.20cm} C{2.20cm}  }
\thickhline
Data & method & DSC & HD95 (mm) & ASSD (mm)  \\ \thickhline
Multi-shell, m=270 & TractSeg & $0.829 \pm 0.056$ & $2.50 \pm 1.33$ & $0.740 \pm 0.202$ \\ \hline
\multirow{4}{*}{\parbox{3.0cm}{b=1000 shell, m=90}}  & TractSeg  & $0.800 \pm 0.071$ & $2.78 \pm 1.51$ & $0.799 \pm 0.285$   \\
& MAS-FA  & $0.765 \pm 0.080$ & $3.12 \pm 2.01$ & $1.004 \pm 0.801$   \\
& MAS-FOD & $0.792 \pm 0.076$ & $2.76 \pm 1.55$ & $0.815 \pm 0.289$   \\
& Proposed & $0.826 \pm 0.056^*$ & $2.48 \pm 1.28^*$ & $0.746 \pm 0.201^*$   \\ \hline
\multirow{4}{*}{\parbox{3.0cm}{b=1000 shell, m=6}}  & TractSeg  & $0.687 \pm 0.155$ & $5.29 \pm 6.51$ & $1.471 \pm 1.427$   \\ 
& MAS-FA  & $0.760 \pm 0.089$ & $3.30 \pm 2.27$ & $1.124 \pm 1.038$   \\
& MAS-FOD & $0.693 \pm 0.140$ & $4.10 \pm 3.13$ & $1.270 \pm 1.361$   \\
& Proposed & $0.825 \pm 0.058^*$ & $2.48 \pm 1.27^*$ & $0.747 \pm 0.211^*$   \\
\thickhline
\end{tabular}
\end{table*}

\begin{figure}[!htb]
\centering
\includegraphics[width=1.0\textwidth]{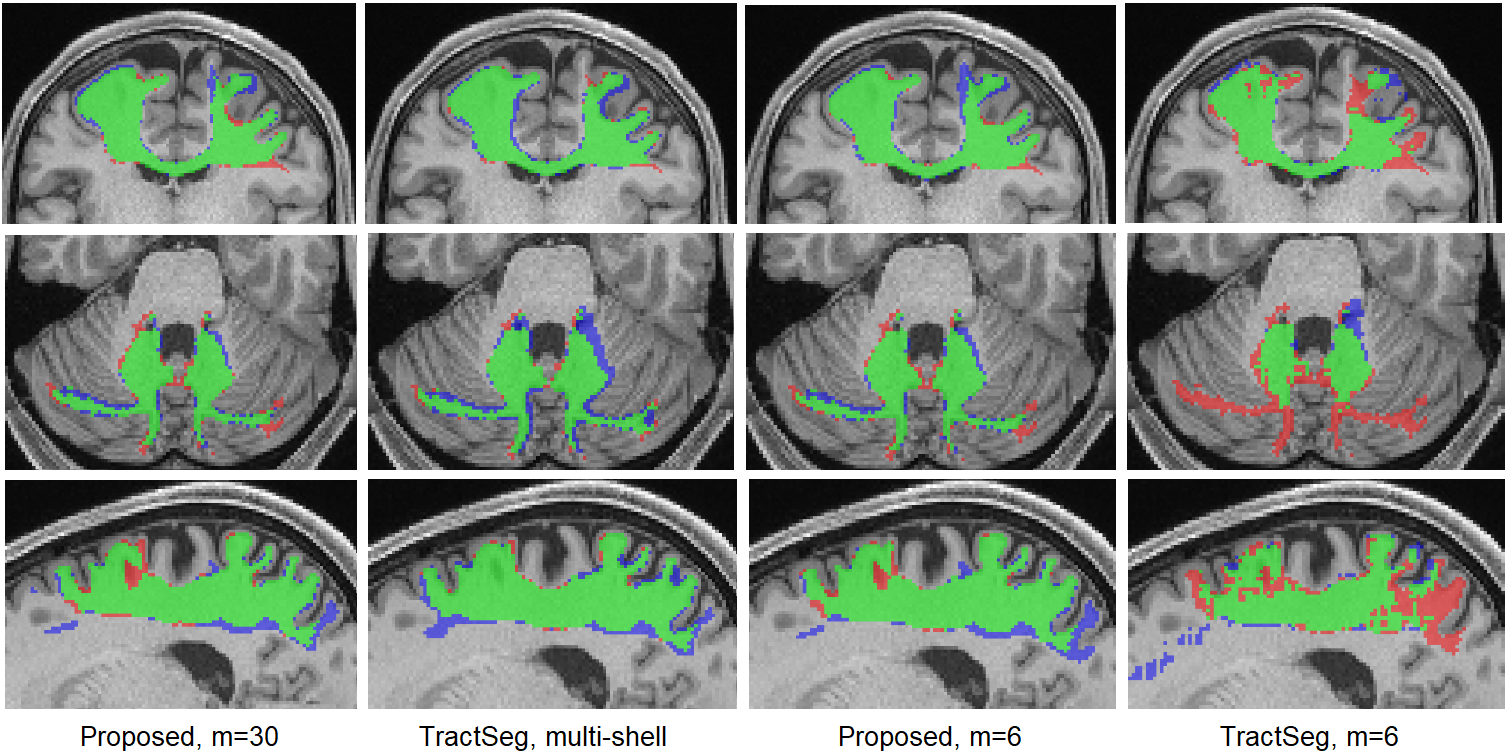}
\caption{\small{Example segmentation results for the proposed method and TractSeg. Green indicates voxels with correct segmentation; red and blue indicate, respectively, false negatives and false positives.}}
\label{fig:segmentation_results}
\end{figure}

Segmentation performance results are presented in Table \ref{table:segmentation_accuracy_table}. Figure \ref{fig:segmentation_results} shows example tract segmentations predicted by our method and TractSeg. Our method using only the DTI measurements (b=1000) achieved segmentation accuracy that was very close to TractSeg using the multi-shell data with three times as many measurements. Paired t-tests did not show any significant differences (at $p=0.01$) between our method and TractSeg in terms of any of the three criteria. When TractSeg was applied on the b=1000 measurements, its performance was worse than our method in terms of all three criteria. To simulate under-sampled clinical scans, we selected 6 of the b=1000 measurements as proposed in \cite{jones1999optimal,skare2000condition}. As shown in Table \ref{table:segmentation_accuracy_table}, the performance of our method remained almost unchanged, whereas the performance of TractSeg deteriorated significantly. Paired t-tests with a $p=0.01$ threshold showed that (1) the performance of our method did not change in terms of any of the three criteria on any of the 41 tracts when 6 measurements were used compared with 90 measurements. (2) Our method achieved significantly higher DSC and significantly lower HD95 and ASSD (all with $p<0.01$) with both 90 and six measurements compared with the other three methods. As shown in Figure \ref{fig:segmentation_results}, segmentations produced by our method are almost indistinguishable between 90 and 6 measurements. Although we cannot present the segmentation results for all tracts, Table \ref{table:segmentation_accuracy_table_tractwise} shows the mean DSC for six of the tracts, including anterior commissure and fornix which were the two most difficult tract to segment for our method and for TractSeg.

\begin{table*}[!htb]
\centering
\small
 \caption{\small{Mean DSC for six tracts. CCg: genu of corpus callosum; ThPr: thalamo-prefrontal; MCP: middle cerebellar peduncle; OpR: optic radiation; AC: anterior commissure; FNX: fornix.  Asterisks denote significantyl better results at $p=0.01$.}}
 \label{table:segmentation_accuracy_table_tractwise}
\begin{tabular}{L{3.0cm}  L{1.5cm} | C{1.1cm} C{1.1cm} C{1.1cm} C{1.1cm} C{1.1cm} C{1.1cm} }
\thickhline
Data & method & CCg  & ThPr & MCP & OpR & AC & FNX  \\ \thickhline
Multi-shell, m=280 & TractSeg & $0.867$ & $0.883$ & $0.871$ & $0.827$ & $0.696$  & $0.689$ \\ \hline
\multirow{2}{*}{\parbox{3.0cm}{Single shell, m=88}}  & TractSeg  & $0.862$ & $0.857$ & $0.826$ & $0.731$ & $0.617$  & $0.528$ \\
& Proposed & $0.901^*$ & $0.897^*$ & $0.864^*$ & $0.810^*$ & $0.703^*$  & $0.675^*$  \\ \hline
\multirow{2}{*}{\parbox{3.0cm}{Single shell, m=6}}  & TractSeg & $0.772$ & $0.783$ & $0.740$ & $0.704$ & $0.366$  & $0.436$  \\ 
& Proposed & $0.903^*$ & $0.897^*$ & $0.857^*$ & $0.811^*$ & $0.680^*$  & $0.666^*$ \\
\thickhline
\end{tabular}
\end{table*}

\begin{figure}[!htb]
\centering
\includegraphics[width=1.0\textwidth]{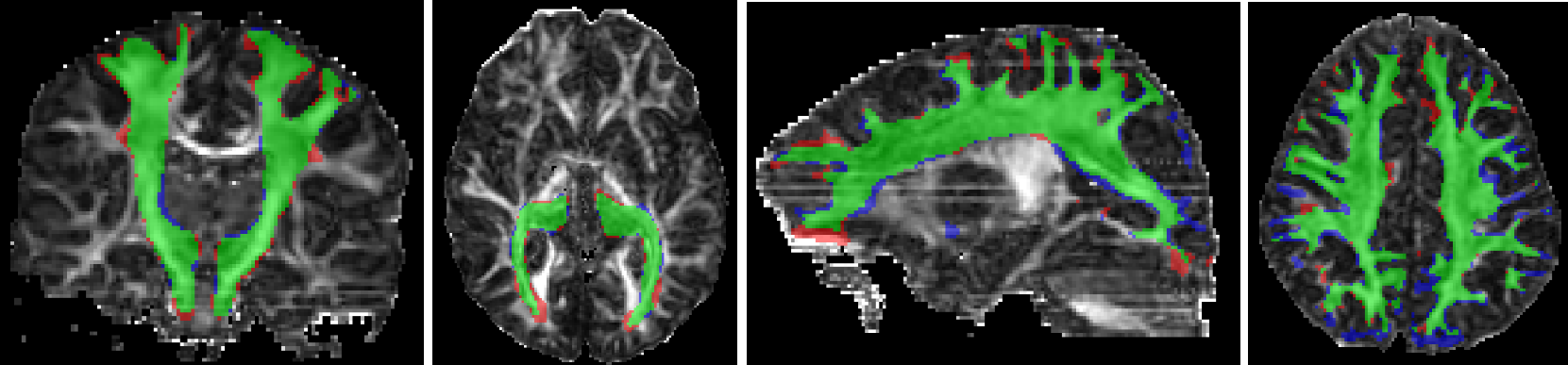}
\caption{\small{Example segmentation results for our proposed method on an independent dataset. Green indicates voxels with correct segmentation; red and blue indicate, respectively, false negatives and false positives.}}
\label{fig:segmentation_results_calgary}
\end{figure}

We further tested our method on scans of children between 2-8 years of age from an independent dataset \cite{reynolds2020calgary}. Each scan in this dataset included 30 measurements in a single shell at b=750. We chose six measurements as input to our model as described above. We manually extracted 32 tracts from 12 different subjects on this dataset. Our method achieved DSC, HD95, and ASSD of $0.786 \pm 0.076$, $2.85 \pm 1.20$, and $1.017 \pm 0.291$, respectively. Although this shows a drop in accuracy, it is a highly encouraging result given the fact that this was a completely independent test dataset that was different from our training dataset in two important ways: (1) subject age: young children (2-8 years) versus adults (21-36 years), and (2) measurement b-value of 750 versus 1000. Compared with our method, TractSeg failed on this dataset, completely missing most of the tracts and achieving a mean DSC of 0.070. To further evaluate the reproducibility of our method on this dataset, we selected two disjoint subsets of six measurements from each scan and applied our method to segment the tracts. We computed the DSC between the tracts computed with the two measurement subsets. We did this for 100 scans, each from a different subject. The DSC for our method was $0.867 \pm 0.041$, whereas it was $0.115 \pm 0.109$ for TractSeg. Example results for our method on this dataset are shown in Figure \ref{fig:segmentation_results_calgary}.

\begin{figure}
\centering
\begin{minipage}{0.55\textwidth}
  \centering
  \includegraphics[width=1.0\linewidth]{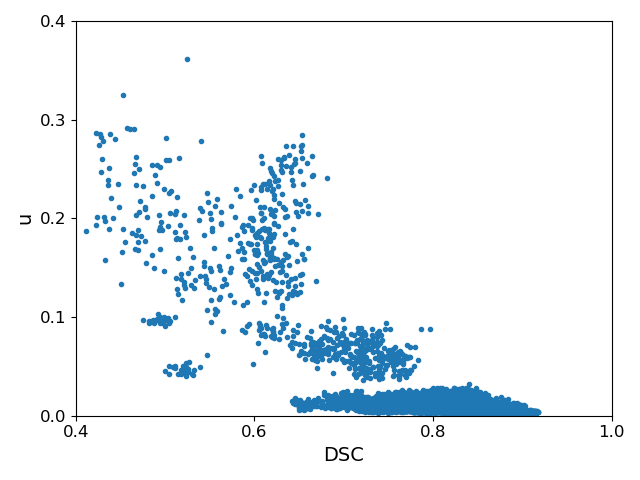}
  \captionof{figure}{Plot of our proposed uncertainty $u$ versus accuracy in terms of DSC.}
  \label{fig:detection}
\end{minipage}%
\begin{minipage}{0.02\textwidth}
$ $
\end{minipage}%
\begin{minipage}{.40\textwidth}
  \centering
    \begin{tabular}{ L{12mm} C{9mm} C{9mm} C{9mm} }
\hline
Method & Acc & Sen & Spc \\
\hline
EMD & 0.91 & 0.86 & 0.92   \\ 
Drp & 0.82 & 0.82 & 0.84   \\ 
Ens & 0.88 & 0.90 & 0.88   \\ 
\hline
\end{tabular}
    \captionof{figure}{Comparison of different methods for identifying inaccurate segmentations, defined as those with DSC<0.70. (Drp: dropout; Ens: ensembles; Acc: accuracy; Sen: sensitivity; Spc: specificity.)}
\label{table:detection_accuracy}
\end{minipage}
\end{figure}

Figure \ref{fig:detection} shows a plot of our proposed segmentation uncertainty, $u$, versus accuracy in terms of DSC. It shows that $u$ is highly effective in identifying the less accurate segmentations. If we choose segmentations with a DSC of 0.70 and lower to be inaccurate, with a threshold of $u=0.30$ we can detect such segmentation with sensitivity=0.86, specificity=0.92, and accuracy=0.91. In Table \ref{table:detection_accuracy} we compare method with the two standard methods based on estimation segmentation uncertainty: dropout, and ensemble methods. We refer to \cite{mehrtash2020confidence} for a description of these methods. Our method achieves overall better results. Note that the ensemble method requires training of multiple models. We trained 10 models in this experiment, which increased the training time by a factor of 10.

\subsection{Computational time and other experiments}

Training time for our method is approximately 24 hours. Our method segments a test image in 2.4 seconds. TractSeg requires approximately 60 seconds to segment an image. MAS methods require much longer time, approximately 3 minutes for MAS-FA and 12 minutes for MAS-FOD.

In recent years attention-based vision models have become very common in medical image segmentation. To experiment with one such model, we applied the model of \cite{karimi2021convolution}, which has been developed specifically for 3D medical image segmentation. This model achieved a DSC of $0.740 \pm 0.125$, which was far lower segmentation performance those reported above.

\section{Conclusions}

Our method shows great promise in segmenting various white matter tracts. The appeal of our method is twofold: (1) Superior accuracy on under-sampled data that are typical of clinical scans, as clearly demonstrated by our results in Figure \ref{fig:segmentation_results} and Tables \ref{table:segmentation_accuracy_table} and \ref{table:segmentation_accuracy_table_tractwise}. (2) Superior generalizability to multi-center data. This was clearly demonstrated in our experiment with an independent validation dataset, with some examples presented in Figure \ref{fig:segmentation_results_calgary}.

\section*{Acknowledgment}

This research was supported in part by the National Institute of Biomedical Imaging and Bioengineering, the National Institute of Neurological Disorders and Stroke, and Eunice Kennedy Shriver National Institute of Child Health and Human Development of the National Institutes of Health (NIH) under award numbers R01HD110772, R01NS128281, R01NS106030, R01EB018988, R01EB031849, R01EB032366, and R01HD109395. This research was also partly supported by NVIDIA Corporation and utilized NVIDIA RTX A6000 and RTX A5000 GPUs. The content of this publication is solely the responsibility of the authors and does not necessarily represent the official views of the NIH or NVIDIA. 
This work was also supported by the Swiss National Science Foundation (project 205321-182602). We acknowledge access to the facilities and expertise of the CIBM Center for Biomedical Imaging, a Swiss research center of excellence founded and supported by Lausanne University Hospital (CHUV), University of Lausanne (UNIL), Ecole polytechnique fédérale de Lausanne (EPFL), University of Geneva (UNIGE) and Geneva University Hospitals (HUG).

\clearpage

\bibliographystyle{splncs04}
\bibliography{davoodreferences}

\end{document}